\documentclass[1p]{elsarticle}
\pdfoutput=1
\usepackage[utf8]{inputenc}
\usepackage[english]{babel}
\usepackage[T1]{fontenc}
\usepackage{amsmath, amssymb}
\usepackage{hyperref}

\usepackage[numbers]{natbib}

\usepackage{graphicx}
\usepackage[hang,small,bf]{caption}
\usepackage[subrefformat=parens]{subcaption}
\usepackage{amsmath, amssymb}
\usepackage{here}
\usepackage{comment}
\usepackage{color}
\usepackage[braket]{qcircuit}
\usepackage{algorithm}
\usepackage{algorithmic}

\usepackage{ulem}

\newcommand{\bx}{\boldsymbol{x}}
\newcommand{\mybv}{\boldsymbol{v}}
\newcommand{\bF}{\boldsymbol{F}}
\newcommand{\dx}{\Delta x}
\newcommand{\mydv}{\Delta v}
\newcommand{\dt}{\Delta t}

\newcommand{\MARK}[1]{{\color{black} #1}}

\DeclareCaptionFormat{algorithm}{\hrulefill\par#1#2#3\par\hrulefill}
\begin{document}

\title{Quantum algorithm for collisionless Boltzmann simulation of self-gravitating systems}

\author[1]{Soichiro Yamazaki}


\author[1,2,3]{Fumio Uchida}


\author[2,4]{Kotaro Fujisawa}


\author[5]{Koichi Miyamoto}

\author[1,6]{Naoki Yoshida}


\affiliation[1]{organization={Department of Physics, The University of Tokyo}, addressline={7-3-1 Hongo}, city={Bunkyo}, postcode={113-0033}, state={Tokyo}, country={Japan}}
\affiliation[2]{organization={Research Center for the Early Universe (RESCEU)}, addressline={7-3-1 Hongo}, city={Bunkyo}, postcode={113-0033}, state={Tokyo}, country={Japan}}
\affiliation[3]{organization={Theory Center, Institute of Particle and Nuclear Studies (IPNS), High Energy Accelerator Research Organization (KEK)}, addressline={1-1 Oho}, city={Tsukuba}, postcode={305-0801}, state={Ibaraki}, country={Japan}}
\affiliation[4]{organization={Department of Liberal Arts, Tokyo University of Technology}, addressline={5-23-22 Nishikamata}, city={Ota}, postcode={144-8535}, state={Tokyo}, country={Japan}}
\affiliation[5]{organization={Center for Quantum Information and Quantum Biology, Osaka University}, addressline={1-2 Machikaneyama}, city={Toyonaka}, postcode={560-0043}, state={Osaka}, country={Japan}}
\affiliation[6]{organization={Kavli Institute for the Physics and Mathematics of the Universe (WPI), UT Institutes for Advanced Study, The University of Tokyo}, city={Kashiwa}, postcode={277-8583}, state={Chiba}, country={Japan}}

\begin{abstract}
The collisionless Boltzmann equation (CBE) is a fundamental equation that governs the dynamics of a broad range of astrophysical systems from space plasma to star clusters and galaxies.
It is computationally expensive to integrate the CBE directly in a multi-dimensional phase space, and thus the applications to realistic astrophysical problems have been limited so far.
Recently, Todorova \& Steijl (2020) proposed an efficient quantum algorithm to solve the CBE with significantly reduced computational complexity.
We extend the algorithm to perform quantum simulations of self-gravitating systems, incorporating the method to calculate gravity with the major Fourier modes of the density distribution extracted from the solution-encoding quantum state.
Our method improves the dependency of time and space complexities on $N_v$, the number of grid points in each velocity coordinate, compared to the classical simulation methods.
We then conduct some numerical demonstrations of our method.
We first run a 1+1 dimensional test calculation of free streaming motion on 64$\times$64 grids using 13 simulated qubits and validate our method. 
We then perform simulations of Jeans collapse, and compare the result with analytic and linear theory calculations.
It will thus allow us to perform large-scale CBE simulations on future quantum computers. 

\end{abstract}
\begin{keyword}
Quantum computing \sep Collisionless Boltzmann equation
\end{keyword}


\maketitle

\section{Introduction}
 A wide variety of numerical simulations are performed in astrophysics to study the formation of galaxies, clusters of galaxies, and the large-scale structure of the universe. Often particle-based $N$-body methods are employed to follow the gravitational dynamics. 
Although there exist well-known problems such as artificial two-body relaxation and shot noise in discrete $N$-body simulations, the particle-based method has been a practical choice since it is computationally less expensive than numerical integration of the collisionless Boltzmann equation (CBE).
A critical problem with direct Boltzmann simulations is the large dimension of the phase space to be considered; a simulation with full three spatial dimensions requires integration of CBE in a six-dimensional phase space. So far, such applications of CBE solvers have been limited even on
supercomputers that are capable of more than $10^{15}$ calculations per second \citep{2013ApJ...762..116Y}.

Quantum computation may hold promise for performing
numerical simulations with large dimensions.
Recently, a novel and efficient scheme was proposed to solve CBE on a quantum computer \citep{2020JCoPh.40909347T}.
It is based on a quantum version of the so-called reservoir method for simulations of hyperbolic systems. Its successful applications include both classical and quantum simulations of hyperbolic systems with conservation laws \citep{Alouges08,Fillion19}.
Ref.~\citep{2020JCoPh.40909347T} performs quantum simulations of
free-molecular flows and flows under a homogeneous force field. It would be highly interesting if the 
quantum computation can also be applied to self-gravitating systems.
Here, we propose an efficient algorithm that can also treat a variable force field. We also propose novel methods for generating initial quantum states and for retrieving information from intermediate and final results of a simulation.
With these implementations, we perform a set of test calculations and validate our numerical scheme.

The rest of the paper is organized as follows.
We introduce our computational scheme in Sec.~\ref{sec2}, and then explain our methods to generate initial quantum states and to retrieve information from qubit arrays in Sec.~\ref{sec3}.
We then discuss the computational complexity of the algorithm in Sec.~\ref{sec4}, and finally show the results of simulations of self-gravitating systems in Sec.~\ref{sec5}.

\section{\label{computationalscheme} Computational Scheme}
\label{sec2}
\subsection{Collisionless Boltzmann equation}
The Boltzmann equation is a kinetic equation that describes the behavior of a system with a large number of particles, and is written  
\begin{equation}
    \frac{\partial f}{\partial t} + \mybv \cdot \frac{\partial f}{\partial \bx} + \bF \cdot \frac{\partial f}{\partial \mybv} = C(f,f')
\end{equation}
where $\bF$ denotes the force field per unit mass and the right-hand side expresses the collision term. 
In this paper, we consider $\bF$ as self-gravity.
The velocity distribution function $f(\bx,\mybv,t)$ represents the fraction of
particles existing in a small phase space volume ${\rm d}\bx \,{\rm d}\mybv$. 
Throughout the present paper,
we consider collisionless
systems with $C(f,f')=0$, which represent, for instance, self-gravitating stars and dark-matter systems in astrophysics. 

\subsection{The reservoir method}
Let us consider a one-dimensional collisionless system. The governing equation is
\begin{equation}
    \frac{\partial f}{\partial t} + v \frac{\partial f}{\partial x} + F 
    \frac{\partial f}{\partial v} = 0.
\end{equation}
We consider operator splitting as
\begin{equation}
    \frac{\partial f}{\partial t} + v \frac{\partial f}{\partial x} = 0,
\end{equation}
\begin{equation}
    \frac{\partial f}{\partial t} + F \frac{\partial f}{\partial v} = 0,
\end{equation}
which can then be discretized in time $\dt_n$ and in phase space $\dx \times \mydv$
using upwind differencing:
\begin{equation}
    f_{k;j}^{n+1} = f_{k;j}^{n} - v_k\frac{\dt_n}{\dx}
    \begin{cases}
        f_{k;j}^{n} - f_{k;j-1}^{n} & (v_k > 0) \\
        f_{k;j+1}^{n} - f_{k;j}^{n} & (v_k < 0)
    \end{cases}
\end{equation}
\begin{equation}
    f_{k;j}^{n+1} = f_{k;j}^{n} - F_j\frac{\dt_n}{\mydv}
    \begin{cases}
        f_{k;j}^{n} - f_{k-1;j}^{n} & (F_j > 0) \\
        f_{k+1;j}^{n} - f_{k;j}^{n} & (F_j < 0)
    \end{cases}
\end{equation}
where $f_{k;j}^{n}$ abbreviates $f(t_n, x_j, v_k)$
with
\begin{equation}
    t_n = \sum_{i=0}^{n-1}\dt_i, \,\,\,x_j = j\dx, \,\,\,\,v_k = k\mydv.
\end{equation}
The reservoir method \citep{Alouges08} utilizes variables $C_k, D_j$ called CFL counters to 
set the time when $f_{k;j}$ is updated.
The CFL counters are initialized to 0 at the beginning and then
updated over $\dt_n$ as
\begin{equation}
    C_k \leftarrow C_k + v_k\frac{\dt_n}{\dx}
\end{equation}
\begin{equation}
    D_j \leftarrow D_j + F_j\frac{\dt_n}{\mydv},
\end{equation}
where $\dt_n$ is chosen such that $|C_k|$ and $|D_j|$ do not exceed unity, 
to satisfy the CFL criteria. 
The time-stepping proceeds such that $f_{k;j}$ is updated to $f_{k;j\mp1}$ when $C_k = \pm1$, and when $D_j = \pm1$, $f_{k;j}$ is updated to $f_{k\mp1;j}$. After each update, we reset the CFL counters whose absolute values are 1. This is the essence of the reservoir method.

For the higher-dimensional case with the $(d+d)$ dimensional phase space, we can extend the above method straightforwardly: we just consider that the indices $k$ and $j$ represent the grid points in the $d$ dimensional configuration and velocity space, respectively.
If we take the same $N_v$ grid points in each velocity coordinate, the number of the CFL counters $C_k$ is effectively $N_v$: for the velocities with equal absolute value and different directions, the CFL counters reach $\pm 1$ simultaneously.
On the other hand, in the general case that the force field $\bF$ depends on $\bx$, $D_j$'s are introduced for each configuration grid point, and thus the number of them is $N_x^d$ in total, where $N_x$ is the number of the grid points in one configuration coordinate. 

We note that there is manifestly no dissipation in this method in the sense that the value of $f_{k;j}$ does {\it not} vary over time, but it is simply advected in phase space. This feature makes the application of the reservoir method to quantum computing as an excellent choice.

\section{Quantum Algorithm}
\label{sec3}

\begin{algorithm}[H]
\caption{Collisionless Boltzmann simulation}
\textbf{Input:} Initial state $f^0_{k;j}$, the number of Fourier modes to be extracted $S$, the number of the time steps $N_t$.
\begin{algorithmic}[1]
\FOR{$n=0$ to $N_t-1$}
\STATE   Construct the initial quantum state $\ket{f^0}$ by QRAM storing $f^0_{k;j}$ on the system $R$ consisting of the registers $R_x$ and $R_v$.
\FORALL{$t\in \{t_i\}$ such that $t\leq n\Delta x/\max(|v_k|)$}
\IF{$t=l\Delta x/\max(|v_k|)$ with some integer $l$}
\STATE   Perform advection in velocity space.
\ENDIF
\STATE   Perform advection in configuration space.
\ENDFOR
\STATE   From the resulting quantum state $\ket{f^n}$, extract $S$ lower-order Fourier modes.
\STATE   Calculate self-gravity \MARK{by solving the Poison equation using the extracted Fourier modes,} and store it in the classical memory.
\ENDFOR
\end{algorithmic}
\label{alg:main}
\end{algorithm}

We repeat four steps to perform quantum Boltzmann simulation using the reservoir method: 
     (1) generating initial conditions in the form of quantum states,
     (2) updating $f_{k;j}$ through a sequence of quantum gate operations,
     (3) extracting information from the quantum states,
     and
     (4) calculating self-gravity.
We give the detailed description for this procedure as Algorithm \ref{alg:main}.
We explain each of the steps (1)-(4) as follows.
We note that we consider the $(1+1)$-dimensional phase space for simplicity. The algorithm can be extended to higher-dimensional cases in a straightforward way.

We hereafter assume periodic boundary conditions in the configuration space.

\subsection{Initialization}
\label{init}
Suppose that a phase space volume in 1+1 dimension is divided into $N_x\times N_v$ grids, where $N_x = 2^{n_x}$ and $N_v = 2^{n_v}$.
In addition to ancillary qubits, a total of $n_x+n_v$ qubits are needed to represent the velocity distribution function by quantum states and manipulate them.
$n_x$ qubits constitute the register $R_x$ that represents the position coordinate $x$, and $n_v$ qubits constitute the register $R_v$ that represents the velocity coordinate $v$.
In the following, we redefine $x_j=j$ and $v_k=(2k+1)\frac{V}{2^{n_v}}-V$, where $V$ is the maximum velocity assumed in our simulations.
      
\begin{figure*}[htbp]
            \begin{minipage}[b]{0.5\linewidth}
                \centering
                \[
\Qcircuit @C=1em @R=1em {
    \lstick{} & \qw & \multigate{3}{MX(j)} & \ctrl{1}  & \multigate{3}{MX(j)} & \qw \\
    \lstick{} & \qw & \ghost{MX(j)} & \ctrl{2} & \ghost{MX(j)} & \qw \\
    &\vdots & \nghost{MX(j)} & & & \\
    \lstick{} & \qw & \ghost{MX(j)} & \ctrl{1} & \ghost{MX(j)} & \qw
\inputgroupv{1}{4}{.5em}{2.9em}{R_x}\\
    \lstick{} & \qw & \qw & \multigate{3}{\parbox{4.6em}{Increment by $\lfloor D_j \rfloor$}} & \qw & \qw \\
    \lstick{} & \qw & \qw & \ghost{\parbox{4.6em}{Increment by $\lfloor D_j \rfloor$}} & \qw & \qw \\
    & \vdots &  & \nghost{\parbox{4.6em}{Increment by $\lfloor D_j \rfloor$}} & & \\
    \lstick{} & \qw & \qw & \ghost{\parbox{4.6em}{Increment by $\lfloor D_j \rfloor$}} & \qw & \qw
\inputgroupv{5}{8}{.5em}{2.8em}{R_v}
}
\]
                \subcaption{$D_j > 0$}
                \label{speed>}
            \end{minipage}
            \hspace{0.06\columnwidth}
            \begin{minipage}[b]{0.5\linewidth}
                \centering
                \[
\Qcircuit @C=1em @R=1em {
    \lstick{} & \qw & \multigate{3}{MX(j)} & \ctrl{1}  & \multigate{3}{MX(j)} & \qw \\
    \lstick{} & \qw & \ghost{MX(j)} & \ctrl{2} & \ghost{MX(j)} & \qw \\
    &\vdots & \nghost{MX(j)} & & & \\
    \lstick{} & \qw & \ghost{MX(j)} & \ctrl{1} & \ghost{MX(j)} & \qw
\inputgroupv{1}{4}{.5em}{2.9em}{R_x}\\
    \lstick{} & \qw & \qw & \multigate{3}{\parbox{4.6em}{Decrement by $\lceil D_j \rceil$}} & \qw & \qw \\
    \lstick{} & \qw & \qw & \ghost{\parbox{4.6em}{Decrement by $\lceil D_j \rceil$}} & \qw & \qw \\
    & \vdots &  & \nghost{\parbox{4.6em}{Decrement by $\lceil D_j \rceil$}} & & \\
    \lstick{} & \qw & \qw & \ghost{\parbox{4.6em}{Decrement by $\lceil D_j \rceil$}} & \qw & \qw
\inputgroupv{5}{8}{.5em}{2.8em}{R_v}
}
\]
                \subcaption{$D_j < 0$}
                \label{speed<}
            \end{minipage}
            \caption{Quantum circuits for advection in velocity space.
             A quantum register assigned to the space (velocity) coordinate is denoted by $R_x$ ($R_v$).
            The component $MX(j)$ is an array of $X$ gates, by which the increment or decrement is activated only when $R_x$ takes $\ket{j}$.
            Details of $MX(j)$ and increment/decrement are explained in Appendix \ref{ap:quantum_gates}.}
            \label{quantum_gates}
\end{figure*}
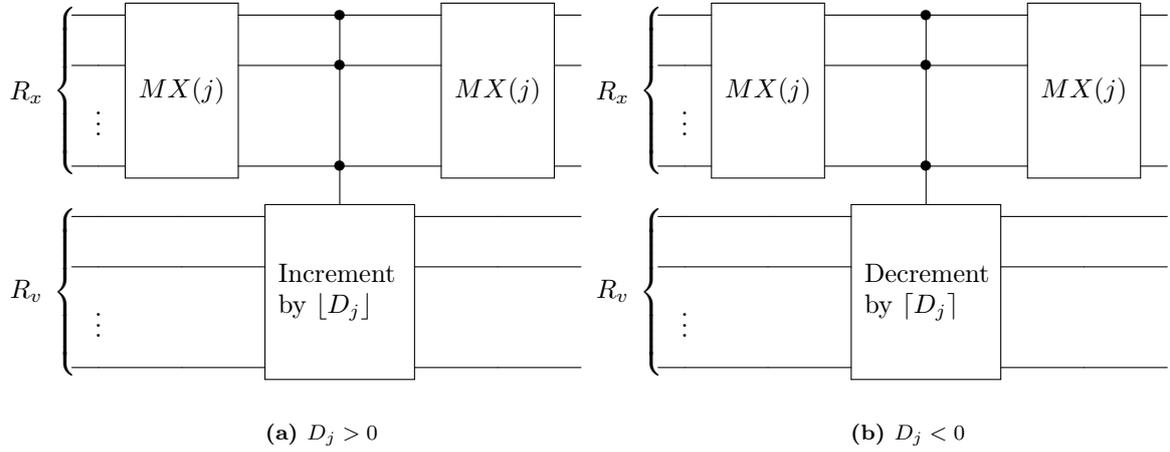
        
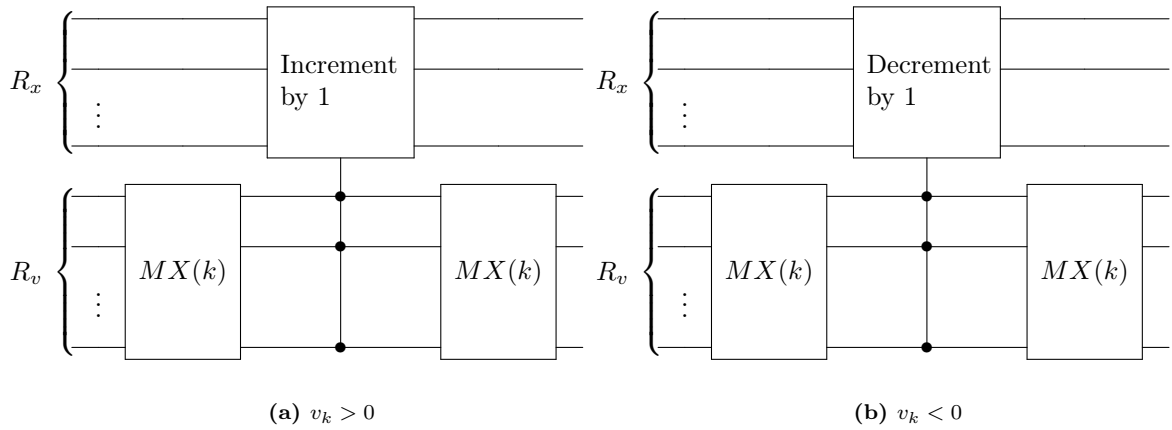
\begin{figure*}[htbp]
            \begin{minipage}[b]{0.5\linewidth}
                \centering
                \[
\Qcircuit @C=1em @R=1em {
    \lstick{} & \qw & \qw & \multigate{3}{\parbox{4.5em}{Increment by 1}} & \qw & \qw \\
    \lstick{} & \qw & \qw & \ghost{\parbox{4.5em}{Increment by 1}} & \qw & \qw \\
     & \vdots &&&&  \\
    \lstick{} & \qw & \qw & \ghost{\parbox{4.5em}{Increment by 1}} & \qw & \qw
\inputgroupv{1}{4}{.5em}{2.4em}{R_x}\\
    \lstick{} & \qw & \multigate{3}{MX(k)} & \ctrl{-1}  & \multigate{3}{MX(k)} & \qw \\
    \lstick{} & \qw & \ghost{MX(k)} & \ctrl{-1} & \ghost{MX(k)} & \qw \\
    \lstick{} & \vdots & \nghost{MX(k)} & & \nghost{MX(k)} & \\
    \lstick{} & \qw & \ghost{MX(k)} & \ctrl{-2}  & \ghost{MX(k)} & \qw
\inputgroupv{5}{8}{.5em}{2.9em}{R_v}
}
\]
                \subcaption{$v_k > 0$}
                \label{space>}
            \end{minipage}
            \hspace{0.06\columnwidth}
            \begin{minipage}[b]{0.5\linewidth}
                \centering
                \[
\Qcircuit @C=1em @R=1em {
    \lstick{} & \qw & \qw & \multigate{3}{\parbox{4.5em}{Decrement by 1}} & \qw & \qw \\
    \lstick{} & \qw & \qw & \ghost{\parbox{4.5em}{Decrement by 1}} & \qw & \qw \\
     & \vdots &&&&  \\
    \lstick{} & \qw & \qw & \ghost{\parbox{4.5em}{Decrement by 1}} & \qw & \qw
\inputgroupv{1}{4}{.5em}{2.4em}{R_x}\\
    \lstick{} & \qw & \multigate{3}{MX(k)} & \ctrl{-1}  & \multigate{3}{MX(k)} & \qw \\
    \lstick{} & \qw & \ghost{MX(k)} & \ctrl{-1} & \ghost{MX(k)} & \qw \\
    \lstick{} & \vdots & \nghost{MX(k)} & & \nghost{MX(k)} & \\
    \lstick{} & \qw & \ghost{MX(k)} & \ctrl{-2}  & \ghost{MX(k)} & \qw
\inputgroupv{5}{8}{.5em}{2.9em}{R_v}
}
\]
                \subcaption{$v_k < 0$}
                \label{space<}
            \end{minipage}
            \caption{Quantum circuits for advection in configuration space.
            }
\end{figure*}
        
We first need to encode the initial distribution $f^0_{k;j}$ into the amplitudes of a quantum state.
Namely, we generate the following quantum state:
\begin{eqnarray}
    \ket{f^0}
    &=& \frac{1}{M}\sum_{j=0}^{2^{n_x}-1}\sum_{k=0}^{2^{n_v}-1}f^0_{k;j}\ket{j}\ket{k}, \nonumber \\ \label{eq:initial_state}
    &&\quad\left(M^2 = \sum_{j=0}^{N_x-1}\sum_{k=0}^{N_v-1}f_{k;j}^2\right).
\end{eqnarray}
To this end, we assume that a QRAM is available \cite{PhysRevLett.100.160501, 2008PhRvA..78e2310G}.
QRAM is a quantum device that can embed classical binary data $x_i$ into the quantum state
\begin{equation}
    \ket{i}\ket{0} \xrightarrow{\mbox{QRAM}} \ket{i}\ket{x_i}.
\end{equation}
Given this, we can use the amplitude encoding technique in \cite{2014Prakash} to generate the following quantum state
\begin{equation}
    \frac{1}{M}\sum_{i=0}^{2^{n_a}-1}x_i\ket{i}_a,\qquad M^2=\sum_{i=0}^{2^{n_a}-1}x_i^2.
\end{equation}
Thus, given a QRAM that stores $f^0_{k;j}$, we can prepare the initial state in Eq.~(\ref{eq:initial_state}).

\subsection{Updating $f_{k;j}$}
From the definition of $v_k$, the time when $C_k = \pm 1$ is calculated as 
\begin{equation}
    \begin{split}
        &\{t_i\}_{i=0,1,2,\dots} = \mbox{sorted sequence of} \\ &\left\{\frac{i}{\vert v_k\vert}~\middle|~i=0,1,2,\dots,~~k=0,1,\dots,2^{n_v}-1\right\}.
    \end{split}
\end{equation}
We fix the discrete time points in the simulation to these points.



\subsubsection{Advection in velocity space}
This set of operations is performed at time $t$ such that $t=l\Delta x/\max(|v_k|)$ with some integer $l$.
We perform the following operations for each $j=0,\ldots,N_v-1$.
First, we update the CFL counter $D_j$ as
\begin{equation}
    D_j \leftarrow D_j + F_j\frac{\Delta x}{\max(|v_k|) \mydv}.
    \label{eq:DjUpdate}
\end{equation}
If $D_j > 0$, we update $f_{k;j}$ to $f_{k-\lfloor D_j \rfloor;j}$, which corresponds to the $\lfloor D_j \rfloor$-time advection in velocity space.
To perform this on $f_{k;j}$ encoded in the amplitudes of the quantum state, we operate the quantum circuit in Fig.~\ref{speed>}.
We then reset $D_j$ as
\begin{equation}
    D_j \leftarrow D_j - \lfloor D_j\rfloor.
\end{equation}
If $D_j < 0$ otherwise, we use a quantum circuit in Fig.~\ref{speed<} to update $f_{k;j}$ to $f_{k+\lceil D_j \rceil;j}$ in the quantum state, and reset $D_j$ as
\begin{equation}
    D_j \leftarrow D_j - \lceil D_j\rceil.
\end{equation}
Here, $\lfloor \cdot \rfloor$ is a floor function that returns the largest integer less than or equal to the input number, and $\lceil \cdot \rceil$ is a ceiling function that returns the smallest integer more than or equal to the input number.


Note that because the increment and decrement in Figs.~\ref{speed>} and \ref{speed<} are modulo $N_v$, if $k-\lfloor D_j \rfloor$ or $k+\lceil D_j \rceil$ falls out of the range from 0 to $N_v-1$, it is pushed back to that range by taking the remainder divided by $N_v$, which is an unphysical operation.
We assume that this never occurs in our simulation because the domain $[-V,V]$ in the velocity space is configured to be sufficiently large. 
        
\subsubsection{Advection in configuration space}
At time $t$, we perform the advection operation only for velocity grids with the index $k$ such that $t v_k\in \mathbb{Z}$.
If $v_k > 0$ and $v_k<0$, we operate the quantum circuit in Fig.~\ref{space>} and the one in Fig.~\ref{space<}, respectively.
Here, the increment or decrement by 1 is operated on $R_x$.
Unlike advection in the velocity space, the modular increment and decrement in advection in the configuration space fit the periodic boundary condition and thus cause no problems. 

The above sequence of operations completes the quantum version of the reservoir method in phase space.

\subsection{Information extraction} \label{section:extraction}

After the advection operations, the quantum state is
\begin{equation}
    \ket{f^n} = \frac{1}{M}\sum_{j=0}^{N_x-1}\sum_{k=0}^{N_v-1}f^n_{k;j}\ket{j}\ket{k}.
\end{equation}
By using $H$ gates to the register $R_v$, it becomes
\begin{align}
    &\left(I^{\otimes n_x} \otimes H^{\otimes n_v}\right)\ket{f^n} \nonumber \\
    &= \frac{1}{M}\sum_{j=0}^{N_x-1}\sum_{k=0}^{N_v-1}\sum_{l=0}^{N_v-1}f^n_{k;j}\frac{(-1)^{k\cdot l}}{\sqrt{N_v}}\ket{j}\ket{l} \nonumber \\
    &= \frac{1}{\sqrt{N_v}M}\sum_{j=0}^{N_x-1}\sum_{l=0}^{N_v-1}\left(\sum_{k=0}^{N_v-1}(-1)^{k\cdot l}f^n_{k;j}\right)\ket{j}\ket{l},
    \label{eq:hgate_int}
\end{align}
where $\cdot$ denotes a sum over digits of binary expression of bitwise AND
(e.g. $5\cdot6 = (101)_{(2)}\cdot(110)_{(2)} = 1+0+0 = 1$).
The lowest component ($l=0$) in Eq. (\ref{eq:hgate_int}) yields density
\begin{equation} \label{eq:full_rho}
    \left(\sum_{k=0}^{N_v-1}(-1)^{k\cdot 0}f^n_{k;j}\right)\ket{j}
    = \frac{\rho^n_j}{\Delta v}\ket{j}.
\end{equation}
\MARK{Instead of extracting real-space information of Eq. (\ref{eq:full_rho}), we extract Fourier space information to reduce the number of measurements by abandoning non-robust small-scale information.}
Quantum Fourier transform (QFT) yields
\begin{equation}
    \mathrm{QFT}\left(\sum_{j=0}^{N_x-1} \rho^n_j\ket{j}\right)=  \sum_{k=0}^{N_x-1}\tilde{\rho}^n_k\ket{k},
\end{equation}
where $\tilde{\rho}^n_k:=\frac{1}{\sqrt{N_x}}\sum_{j=0}^{N_x-1}\exp\left(\frac{2\pi ikj}{N_x}\right)\rho^n_j$ is the discrete Fourier transform of $\rho^n_j$.

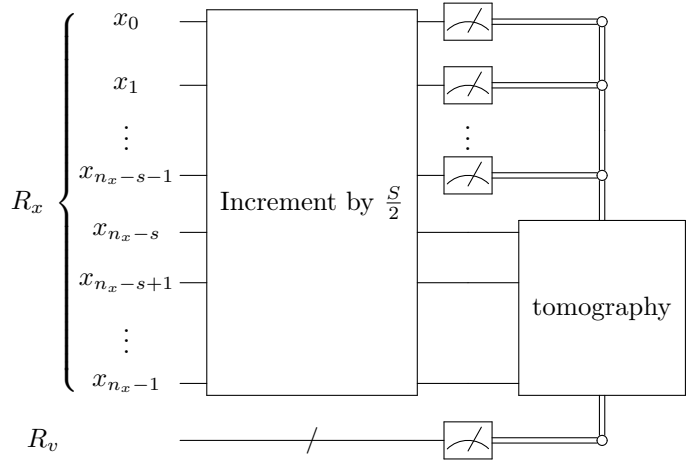
\begin{figure*}
    \centering
    \begin{minipage}[b]{1\linewidth}
    \Qcircuit @C=1em @R=1em {
        \lstick{} && x_0         &&& \multigate{7}{\text{Increment by}~\frac{S}{2}}        & \meter & \controlo \cw \\
        \lstick{} && x_1         &&& \ghost{\text{Increment by}~\frac{S}{2}}        & \meter &  \controlo \cw \cwx \\
        \lstick{} && \vdots      &&&            & \vdots  & \cwx \\
        \lstick{} && x_{n_x-s-1} &&& \ghost{\text{Increment by}~\frac{S}{2}}        & \meter &  \controlo \cw \cwx\\
        \lstick{} && x_{n_x-s}   &&& \ghost{\text{Increment by}~\frac{S}{2}}  & \qw  & \multigate{3}{\mathrm{tomography}} \cwx \\
        \lstick{} && x_{n_x-s+1} &&& \ghost{\text{Increment by}~\frac{S}{2}}    & \qw        & \ghost{\mathrm{tomography}} \\
        \lstick{} && \vdots      &&&                    &    & \nghost{\mathrm{tomography}} \\
        \lstick{} && x_{n_x-1}   &&&   \ghost{\text{Increment by}~\frac{S}{2}}      & \qw  &  \ghost{\mathrm{tomography}}           \inputgroupv{1}{8}{.5em}{6.8em}{R_x} \\
        \lstick{R_v}          &&    &&& {/} \qw           & \meter & \controlo \cw \cwx
    }
    \end{minipage}
    \caption{A quantum circuit for extracting information from the quantum state.}
    \label{fig:before_tom}
\end{figure*}

At this time, we set a new integer parameter $S=2^s$, which determines the number of the Fourier components we extract from the quantum state.
It should meet the conditions below.
\begin{equation}
    1 \leq s \leq n_x-1.
\end{equation}
The quantum circuit used to extract the Fourier components is shown in Fig. \ref{fig:before_tom}.
After an increment of $\frac{S}{2}$ on $R_x$, we measure the first $n_x-s$ qubits in $R_x$ and all the qubits in $R_v$ and postselect the states that all of them are $\ket{0}$, to obtain the state
\begin{equation}
    \ket{\tilde{\rho}^n}:=\frac{1}{\tilde{M}}\left(\sum_{k=0}^{\frac{S}{2}-1} \tilde{\rho}^n_{N_x-\frac{S}{2}+k}\ket{k}+\sum_{k=0}^{\frac{S}{2}-1} \tilde{\rho}^n_{k}\ket{k+\frac{S}{2}}\right),
\end{equation}
where $\tilde{M}^2:=\sum_{k=0}^{\frac{S}{2}-1} \left|\tilde{\rho}^n_{N_x-\frac{S}{2}+k}\right|^2+\sum_{k=0}^{\frac{S}{2}-1} \left|\tilde{\rho}^n_{k}\right|^2$.
This state encodes $\tilde{\rho}^n_{N_x-\frac{S}{2}},$$\ldots,$$\tilde{\rho}^n_{N_x-1},$$\tilde{\rho}^n_{0},$ $\ldots,$$\tilde{\rho}^n_{\frac{S}{2}-1}$, which are the Fourier components of the density with wavenumber $-\frac{S}{2},$$\ldots,$$\frac{S}{2}-1$, respectively.
Then, we extract these large-scale Fourier modes from $\ket{\tilde{\rho}}$ using quantum state tomography, whose detail is described in Appendix \ref{sec:q_tom}.
Considering the system whose large-scale structure is important in its simulation (e.g., neutrino distribution in the large-scale structure of the universe~\cite{Yoshikawa_2020}), the large-wavenumber modes become almost 0, and we can describe the system well only with the large-scale modes.

\subsection{Calculation of the self-gravity}
Self-gravity $\bF$ obeys the following equations:
\begin{equation}
    \bF = -\frac{\partial \phi}{\partial \bx},
\end{equation}
\begin{equation}
    \Delta\phi = 4\pi G\rho,
\end{equation}
where $\phi$ is the gravitational potential.
Then,
\begin{equation}
    \tilde{\phi}(\boldsymbol{k}) = \tilde{G}(\boldsymbol{k})\tilde{\rho}(\boldsymbol{k}),
\end{equation}
where, in the case of 3 dimensions \cite{2013ApJ...762..116Y},
\begin{equation}
    \tilde{G}(\boldsymbol{k}) = -\frac{\pi G}{\sin^2(\frac{\pi k_x}{N_x}) + \sin^2(\frac{\pi k_y}{N_x}) + \sin^2(\frac{\pi k_z}{N_x})},
\end{equation}
and, in the case of 1 dimension,
\begin{equation}
    \tilde{G}(\boldsymbol{k}) = -\frac{\pi G}{\sin^2(\frac{\pi k}{N_x})}.
\end{equation}

\section{\label{computationalcomplexity} Computational complexity}
\label{sec4}
We first discuss the time complexity of our numerical algorithm in the $d$-dimensional case. 
The characteristic time of the system is $T=\Delta x/\mbox{max}(|v_k|)\simeq 1/V$. 
We set the terminal simulation time to $N_tT$ with $N_t\in\mathbb{N}$.
The general accuracy of the solution will not be significantly affected if we calculate $F$ every $T$, and thus we do so.
A multi-controlled NOT gate with $n$ control qubits is assumed to be executed with the time complexity of $\mathcal{O}(n)$ \cite{2013PhRvA..87f2318S}.

Then, let us evaluate the time complexity from advection in velocity space.
As described in Appendix~\ref{ap:quantum_gates}, the circuit for the increment or decrement by $p$ on an $n$-qubit register is implemented with $O(n\log p)$ multi-controlled NOT gates with at most $n$ control qubits, and hence the quantum circuits shown in Fig.~\ref{quantum_gates} cost $\mathcal{O}\left(n_v(n_x+n_v)\log|D_j|\right)$ time.
To extract the small-wavenumber Fourier modes of density at time $t=nT$ with accuracy $\epsilon$ from the quantum state $\ket{\tilde{\rho}}$, quantum tomography forces us to prepare $O(S^d/\epsilon^2)$ copies of the state $\ket{\tilde{\rho}^n}$, where we denote by $S^d$ the number of Fourier modes we extract in the case with $d$ dimensions.
To prepare $\ket{f^n}$ and then $\ket{\tilde{\rho}^n}$, the advection operation is performed $\mathcal{O}(n)$ times.
Because, in estimating the Fourier modes of the density at each time step, we need to start the time evolution from the initial time $\ket{f^0}$, the total number of advection operations in the velocity space to obtain the Fourier modes at the terminal time $N_tT$ is
\begin{equation}
    \mathcal{O}\left(\frac{N_x^dN_t^2S^d}{\epsilon^2}\right),
\end{equation}
where $N_x^d$ is the number of the CFL counters $D_j$.
Post-selection does not change the order of the complexity (see Sec.~\ref{sec5} for the detail).
Consequently, the time complexity for advection in velocity space becomes
\begin{equation}
    \mathcal{O}\left(\frac{n_v(n_x+n_v)N_x^dN_t^2S^d}{\epsilon^2}\log \left(\frac{F_sN_v\Delta x}{V^2}\right)\right),
    \label{eq:compAdVelQ}
\end{equation}
where, according to Eq.~\eqref{eq:DjUpdate}, we evalauate $|D_j|$ as $F_sN_v\Delta x/V^2$ with a characteristic magnitude $F_s$ of $F$.

The time complexity for advection in configuration space is evaluated similarly.
Considering that the number of time points $t_i$ in 
the system's time over $T$ is $\mathcal{O}(N_v)$, 
we evaluate the time complexity for advection in configuration space as
\begin{equation}
    \mathcal{O}\left(\frac{n_x(n_x+n_v)N_vN_t^2S^d}{\epsilon^2}\right).
\end{equation}

\MARK{Compared to the operations for advection, other operations make smaller contributions to the complexity.
As for the QRAM, which we use to generate the initial state $\ket{f^0}$, since $\ket{f^0}$ is generated $\mathcal{O}\left(N_tS^d/\epsilon^2\right)$ times and a query to the QRAM storing $N_x^d N_v^d$ entries takes $\mathcal{O}\left(\log (N_x^d N_v^d)\right)=\mathcal{O}\left(d (n_x+n_v)\right)$ time, the total time complexity concerning the QRAM is
\begin{equation}
    \mathcal{O}\left(\frac{d(n_x+n_v)N_tS^d}{\epsilon^2}\right).
\end{equation}
Besides, the measurements for information extraction described in Sec. \ref{section:extraction} take
\begin{equation}
    \mathcal{O}\left(\frac{N_tS^d}{\epsilon^2}\right)
\end{equation}
times, since the number of the measurements is of this order and the time for one measurement is supposed to be $O(1)$.}

In classical computing, advection in configuration space and that in velocity space both cost
\begin{equation}
    \mathcal{O}(N_x^d N_v^d N_t)
    \label{eq:compCl}
\end{equation}
time complexity.
The latter involves the calculation of gravity, which takes a time complexity of the same order as Eq. \eqref{eq:compCl}~\footnote{
The main contribution comes from calculating the density $\rho^n_j$ by integrating $f^n_{k;j}$.
This is followed by the Fourier transform of the density and the inverse Fourier transform of $\tilde{\phi}$, and these steps amount to $\mathcal{O}(n_x N_x^d N_t)$ time complexity, which is subdominant as far as $n_x \ll N_v^d$.
Although these (inverse) Fourier transforms are done also in the quantum method, their contribution to the time complexity is again subdominant compared to the total complexity in Eq. \eqref{eq:compAdVelQ}.
}.
Compared to Eq. \eqref{eq:compCl}, the quantum method significantly improves the scaling of the time complexity in $N_v$, while the scaling on $N_t$ worsens and the dependence on $\epsilon$ and $S$ appears.



The improvement in space complexity is also achieved.
Classically, the $\mathcal{O}(N_x^dN_v^d)$ memory space is needed to store the values of $f_{k;j}$ at grid points, while the CFL counters and the Fourier modes of the density require the smaller ones of order $\mathcal{O}(N_x^d+N_v+S^d)$ in total.
In our quantum method, the number of qubits needed to run the quantum circuits is only of order $O(n_x+n_v)$, and the classical memory space of order $\mathcal{O}(N_x^d+N_v+S^dN_t)$ is used for the CFL counters and the Fourier modes of density.
Note that in the quantum setting, we need to start from the initial state $\ket{f^0}$ to generate $\ket{f^n}$ for time $nT$ and thus store the Fourier modes at intermediate times, resulting in the $\mathcal{O}(S^dN_t)$ memory space.
Nevertheless, we expect that our quantum method achieves a large reduction in space complexity compared to the classical one of the order $\mathcal{O}(N_x^dN_v^d)$.

\section{Numerical Simulations}
\label{sec5}
We have performed the following test calculations \citep{2013ApJ...762..116Y} using Python version 3.8.6 and Rust version 1.70.0 because simulations on currently available quantum computers are still expensive.


\subsection{Free Streaming}
\begin{figure}[htbp]
    \centering
    \includegraphics[width=6cm]{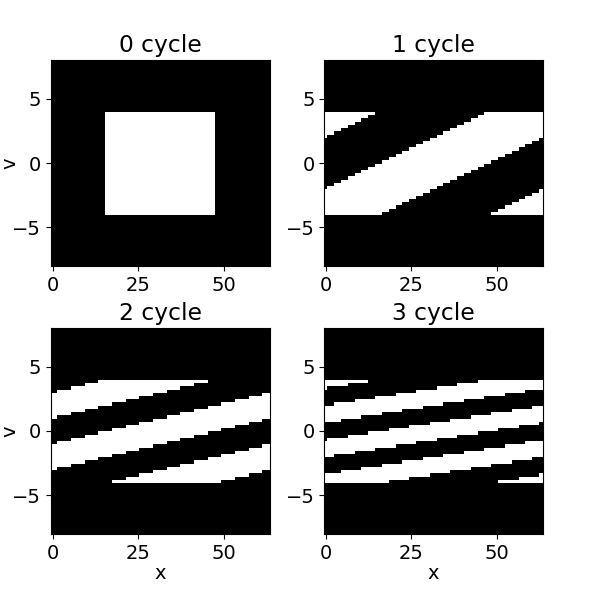}
    \caption{
    Snapshots of the one-dimensional free-streaming simulation with $F=0$ and $n_x=n_v=6$. From left to right, the distribution function $f$ at $t=0, 1, 2$, and $3$ cycles are shown. $1$ cycle is equal to $2\Delta x/\Delta v$. A white "box" at $t=0$ flows without deformation other than shearing in phase space.}
    \label{adv_figs}
\end{figure}
We set $F=0$, $n_x=n_v=6$, and configure a ``box'' initial condition for $f$ as shown in Fig.~\ref{adv_figs}, where the white region has $f=1$ and black region $f=0$.
The exact solution should be simple free streaming in phase space, and $f$ is advected over time to produce
shear motion.
The reservoir method does not produce deformation other than shearing, which we confirm with our numerical result.

\subsection{\label{jeansinstability} Jeans instability}
\begin{figure}[htbp]
    \centering
    \includegraphics[width=8cm]{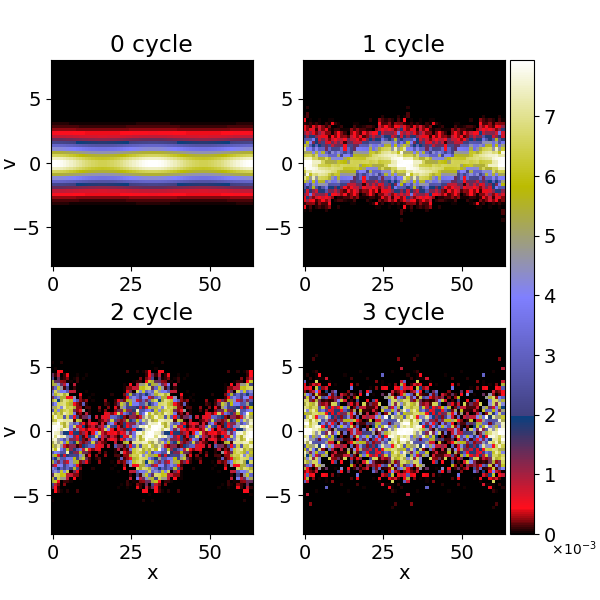}
    \caption{
    One-dimensional Jeans collapse with $n_x=n_v=6$ and with $A=0.1$, $k = 0.5 k_\text{J}$.
    The values of $f$ at $t=0, 1, 2,$ and $3$ cycles are shown
    (see the colorbar on the right).
    }
    \label{adv_figs2}
\end{figure}

\begin{figure}[htbp]
    \centering
    \includegraphics[width=8cm]{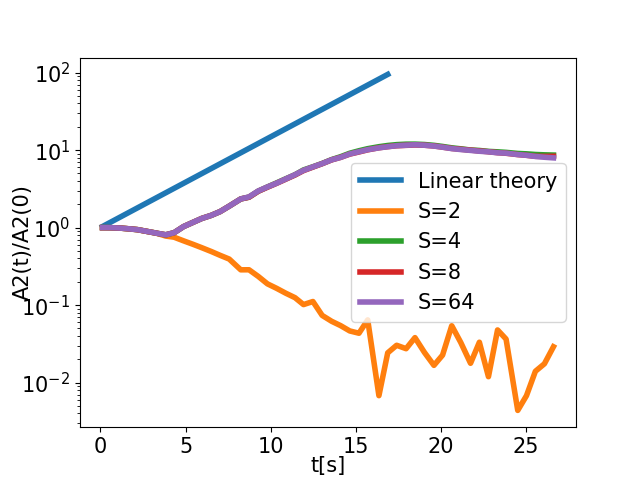}
    \caption{
    The time evolution of the Fourier amplitude 
    $A_2$ of the density fluctuations in the run with  $k/k_\text{J}=0.5$.
    The blue line shows linear growth rate.
    Orange, green, red, and purple lines show the numerical results of $S=2,4,8,64(=N_x)$.
    While the orange line was way off the mark, other numerical results grew at the same speed as linear theory.
    }
    \label{a2_plot}
\end{figure}

Our next example is one-dimensional gravitational instability.
The force field $F$ is the self-gravity of the fluid.
For a homogeneous mass distribution, there exists a stationary solution, i.e., the Maxwell distribution
\begin{equation}
    f_\mathrm{M}(x,v) = \frac{\rho_\mathrm{ref}}{\sqrt{2\pi \sigma^2}}\exp\left(-\frac{v^2}{2\sigma^2}\right).
\end{equation}
We add a perturbation of
\begin{equation}
    \label{bis}
    f(x,v) = f_\mathrm{M}(x,v)(1 + A\cos k x)
\end{equation}
with $A=0.1, k=0.5k_{\text J}$ where the Jeans wavenumber is given by
\begin{equation}
    k_\mathrm{J}=\frac{\sqrt{4\pi G\rho_{\mathrm{ref}}}}{\sigma}.
\end{equation}
Fig.~\ref{adv_figs2} is the result of our simulation.
It shows that the phase-space structure is reproduced accurately up to $t=2$ cycles with the relatively low resolution with
$n_x=n_v=6$, compared to the result of the high-accuracy classical
simulation of Ref.~\citep{2013ApJ...762..116Y} (see their Figure 6 on $128\times 128$ grids). 
At $t=3$ cycles, the distribution appears noisy, and
the spiral structure commonly seen in high-resolution
simulations is not well reproduced. This is likely
owing to the coarse resolution of our simulation.

From the initial condition Eq.~\eqref{bis}, we can calculate the analytical solution for linear growth part of Fourier transform $A_m$ of $\delta(x,t)=\rho(x,t)-\rho_\mathrm{ref}$ \citep{2008gady.book.....B},
which is defined as
\begin{equation}
    \delta (x,t) = \sum_{m \geq 0}A_m(t)\exp(i m\frac{2\pi}{2^{n_x}}x).
\end{equation}
The linear growth rate $\gamma$ is obtained by solving the plasma dispersion relation 
\begin{align}
        (k/k_{\text J})^2 =& 1 + w Z(w), \nonumber \\
         w =& \frac{\pm i\gamma}{\sqrt{8\pi G\rho}\,(k/k_{\text J})}, \nonumber \\
        Z(w) =& \frac{1}{\sqrt{\pi}}\int_{-\infty}^{\infty} {\rm d}s\frac{e^{-s^2}}{s-w}. 
\end{align}
The orange line ($S=2$) in Fig.~\ref{a2_plot} does not show the same growth rate as linear theory (blue line), but other results with $S=4,8,64$ show that the growth of the perturbation is reproduced as predicted by linear theory.
This is due to the double symmetry we set as an initial state, and the orange line, which can use only up to $S-1=1$ st order density information, cannot grasp the existence of the perturbation.
We thus conclude that our numerical scheme is well suited to follow the evolution of one-dimensional self-gravitating systems if $S$ is set large enough.

\subsection{Postselection}

\begin{figure*}[htbp]
    \begin{minipage}[b]{0.45\linewidth}
        \centering
        \includegraphics[width=7cm]{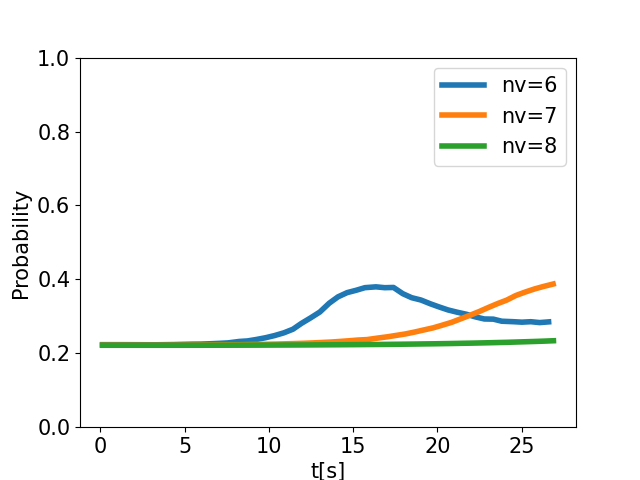}
        \subcaption{Probability that we get 0's on all the qubits in measuring $R_v$. 
    }
        \label{fig:p_h}
    \end{minipage}
    \hspace{0.06\columnwidth}
    \begin{minipage}[b]{0.45\linewidth}
        \centering
        \includegraphics[width=7cm]{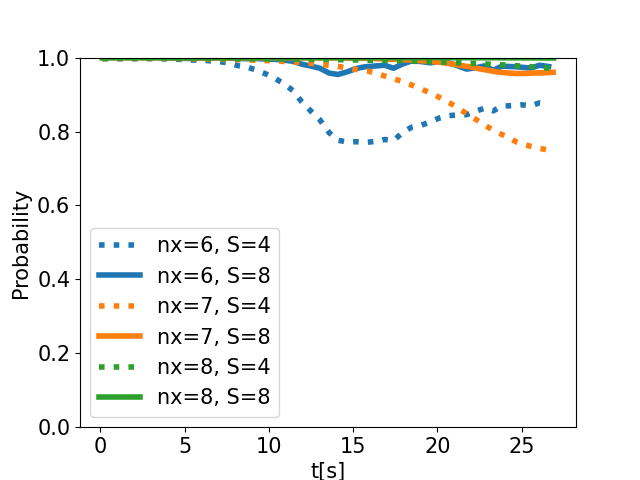}
        \subcaption{Probability that we get 0's on all the first $n_x-s$ qubits in measuring $R_x$, given that the outcomes in measuring the qubits in $R_v$ are all 0.
    }
        \label{fig:m_t}
    \end{minipage}
    \caption{
    Success probability of the post-selection when extracting the large-scale Fourier modes by the circuit in Fig. \ref{fig:before_tom} at each time step in the simulation as for Fig.~\ref{adv_figs2}.
    }
\end{figure*}

In extracting the Fourier modes of density from the quantum state by the circuit in Fig.~\ref{fig:before_tom}, we perform two types of post-selection: we select the states in which all the qubits in $R_v$ take $\ket{0}$, and then select the states in which the first $n_x-s$ qubits in $R_x$ are $\ket{0}$.
We have numerically confirmed that the post-selection does not change the order of the time complexity of our method.
Fig.~\ref{fig:p_h} shows the probability that we get 0's on all the qubits in measuring $R_v$ at each time step in the same simulation as Fig.~\ref{adv_figs2}.
Fig.~\ref{fig:m_t} shows the probability that, after the post-selection of $R_v$, we get 0's on the first $n_x-s$ qubits in measuring $R_x$.
During the entire time shown in Fig.~\ref{fig:p_h} and Fig.~\ref{fig:m_t}, the probabilities stay roughly constant and do not depend on either $n_x$ or $n_v$.
This is because these constants depend on the velocity distribution function rather than on the numerical resolution.
Since we primarily consider physical systems in which small-scale modes are less important, we do not expect significant degradation in computational complexity caused by the post-selection.


\section{Discussion}
\label{sec6}

\begin{figure*}[htbp]
    \begin{minipage}[b]{0.45\linewidth}
        \centering
        \includegraphics[width=7cm]{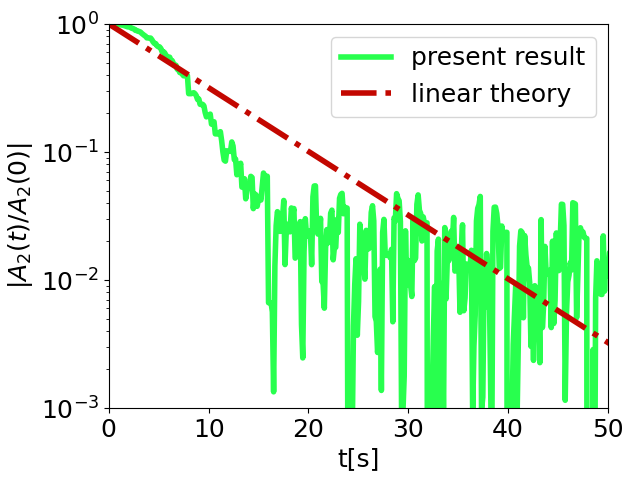}
        \subcaption{$n_v=6$}
        \label{a2_6_6_15}
    \end{minipage}
    \hspace{0.05\columnwidth}
    \begin{minipage}[b]{0.45\linewidth}
        \centering
        \includegraphics[width=7cm]{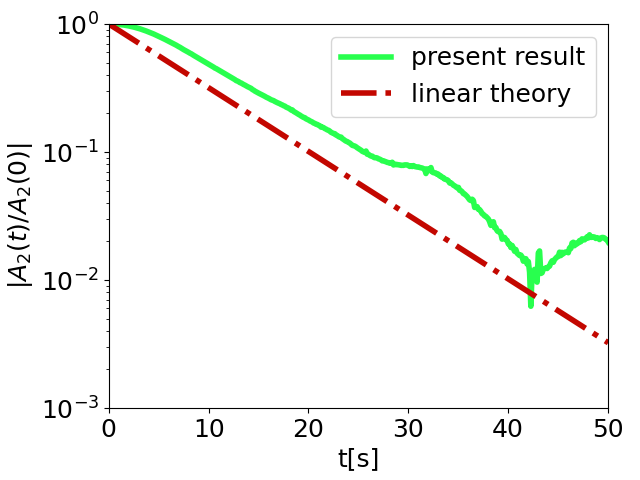}
        \subcaption{$n_v=11$}
        \label{a2_6_11_15}
    \end{minipage}
    \caption{
    Comparison with an analytical solution of Fourier transform $A_2$ of the density fluctuation with $k/k_{\text J}=1.5$.
    The left panel shows the result with $n_v=6$ and the right panel shows the result with $n_v=11$.
    The latter case reproduces $A_2$ with the same growth rate as the analytical solution.
    }
\end{figure*}

    We have developed a quantum algorithm based on the reservoir method to solve the collisionless Boltzmann equation on a quantum computer.
We have conducted classical simulations of our quantum algorithm for some self-gravitating systems, which imply its effectiveness.

We discuss here the limitation of our numerical simulation. 
With the current implementation, the number of discrete grids in velocity space is $N_v$.
The corresponding time step width $\dt_c$ necessary for update is, at each space point,
\begin{equation}
    \dt_c = \frac{\dx}{\max (|v_k|)} = \frac{N_v \dx}{V(N_v-1)},
\end{equation}
whereas the reservoir method requires
\begin{equation}
    F_s \frac{\dt_c}{\mydv} \geq \mathcal{O}(1),
\end{equation}
where $F_s$ is a characteristic magnitude of $F$
and $\mydv = 2V/N_v$.
If this condition is not met, the advection operation in velocity space is not performed for a sufficiently large number of times, which
results in inaccurate evolution of the velocity
distribution function.
This effectively imposes a requirement on the velocity space resolution of
\begin{equation}
    N_v \geq \mathcal{O}\left(\frac{V^2}{F_s\dx}\right).
    \label{eq:resolution}
\end{equation}
To examine the effect of insufficient resolution, we have performed simulations of gravitational Landau damping \citep{2008gady.book.....B} with $k = 1.5 k_{\text J}$ for two setups with $n_v = 6$ and $n_v = 11$. The other conditions are kept the same as in Sec.~\ref{jeansinstability}.
We plot the measured Fourier amplitude of $A_2$ and the damping rate in Fig.~\ref{a2_6_6_15} and \ref{a2_6_11_15}. 
Clearly, the run with poor resolution with $n_v = 6$ yields too rapid damping and spurious oscillations, whereas the run with $n_v = 11$ reproduces the correct initial damping as predicted by linear theory. We find that Eq.~\eqref{eq:resolution} serves as a practical
condition to be checked and monitored through a simulation.

We have proposed a method that utilizes two separate CFL counters for advection in configuration space and in velocity space in an operator-splitting manner (Sec.~\ref{computationalscheme}). 
It allows us to
perform simulations with a time-dependent force field
under the conditions and limitations discussed above.
With the small computational complexity derived in Sec.~\ref{computationalcomplexity}, future quantum computing has the potential to 
perform very large collisionless Boltzmann simulations. 
In our future work, we study applications to general Vlasov--Poisson systems including the neutrino distribution in the large-scale structure of the universe.

\noindent
Funding:
    SY and FU acknowledge financial support from the Forefront Physics and Mathematics Program to Drive Transformation (FoPM).
    KF is supported by JSPS KAKENHI Grant No. JP20K14512.
    KM is supported by MEXT Quantum Leap Flagship Program (MEXT Q-LEAP) Grant no. JPMXS0120319794, JSPS KAKENHI Grant no. JP22K11924, and JST COI-NEXT Program Grant No. JPMJPF2014.


\bibliographystyle{quantum}
\bibliography{yamazaki2022}{}

\appendix

\section{Quantum gates used in the advection circuits}
\label{ap:quantum_gates}

An $n$-qubits gate $MX(N)$, where $0\leq N \leq 2^{n}-1$, is defined by a set of $X$-gates as
\begin{equation}
    MX(N) = \bigotimes_{i=0}^{n-1}X_i^{1-N_i},
\end{equation}
where $N_i$ is the $(n-i)$-th digit of the binary expression of $N$ and $X_i$ is the Pauli X gate on the $i$-th qubit.

\begin{figure}[htbp]
    \centering
    \begin{minipage}[h]{0.45\linewidth}
        \[
        \Qcircuit @C=1em @R=.7em {
            & \targ    & \qw      & \qw      & \qw   & \qw \\
            & \ctrl{-1} & \targ    & \qw      & \qw   & \qw \\
            & \ctrl{-1} & \ctrl{-1} & \targ    & \qw   & \qw \\
            & \ctrl{-1} & \ctrl{-1} & \ctrl{-1} & \targ & \qw
        }
        \]
        \subcaption{Increment by 1}
        \label{fig:increment}
    \end{minipage}
    \begin{minipage}[h]{0.45\linewidth}
        \[
        \Qcircuit @C=1em @R=.7em {
            & \targ    & \qw      & \qw      & \qw   & \qw \\
            & \ctrlo{-1} & \targ    & \qw      & \qw   & \qw \\
            & \ctrlo{-1} & \ctrlo{-1} & \targ    & \qw   & \qw \\
            & \ctrlo{-1} & \ctrlo{-1} & \ctrlo{-1} & \targ & \qw
        }
        \]
        \subcaption{Decrement by 1}
        \label{fig:decrement}
    \end{minipage}
    \caption{Operators for the increment and decrement by 1 in 4-qubit states.}
    \label{fig:inc_dec}
\end{figure}
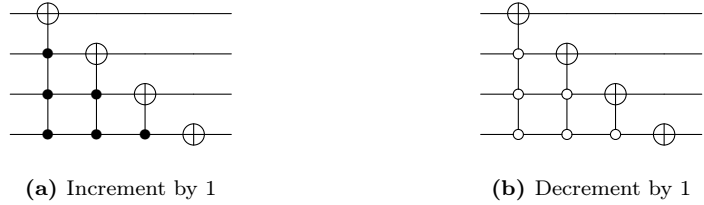

The operators for the increment and decrement by 1 are defined by triangle-like quantum circuits like Fig.~\ref{fig:inc_dec}.
They conduct the following operation
\begin{equation}
    \sum_{i=0}^{2^n-1}a_i |i\rangle \rightarrow \sum_{i=0}^{2^n-1}a_i \ket{i\pm1~{\rm mod}~ 2^n},
\end{equation}
where $+$ represents increment and $-$ represents decrement.

By reducing the size of the triangle, we can realize the addition or subtraction of power of two.
For example, the following 4 qubits operator
\[
        \Qcircuit @C=1em @R=.7em {
            & \targ    & \qw      & \qw      & \qw   & \qw \\
            & \ctrl{-1} & \targ    & \qw      & \qw   & \qw \\
            & \qw & \qw & \qw    & \qw   & \qw \\
            & \qw & \qw & \qw & \qw & \qw
        }
        \]
perform the following operation
\begin{equation}
    \sum_{i=0}^{2^4-1}a_i |i\rangle \rightarrow \sum_{i=0}^{2^4-1}a_i \ket{i+2^2~{\rm mod}~ 2^4},
\end{equation}
namely, the increment by 4.
Combining these operators, we can realize the operator for the increment/decrement by $p$ on an $n$-qubit register, which acts as
\begin{equation}
    \sum_{i=0}^{2^n-1}a_i \ket{i} \rightarrow \sum_{i=0}^{2^n-1}a_i \ket{i+p~{\rm mod}~ 2^n},
\end{equation}
with $\mathcal{O}(n\log p)$ multi-controlled NOT gates.

\section{Quantum state tomography} \label{sec:q_tom}

\begin{figure}[htbp]
    \centering
    \begin{minipage}[h]{0.45\linewidth}
        \Qcircuit @C=1em @R=1em {
            \lstick{} & \meter & \qw \\
            \lstick{} & \meter & \qw \\
            \lstick{} & \meter & \qw \\
            \lstick{} & \meter & \qw
            \inputgroupv{1}{4}{.5em}{3.6em}{\ket{\phi}}
        }
    \end{minipage}
    \hspace{1cm}
    \begin{minipage}[h]{0.45\linewidth}
        \Qcircuit @C=1em @R=1em {
            \lstick{} & \gate{H} & \meter & \qw \\
            \lstick{} & \gate{H} & \meter & \qw \\
            \lstick{} & \gate{H} & \meter & \qw \\
            \lstick{} & \gate{H} & \meter & \qw
            \inputgroupv{1}{4}{.5em}{3.6em}{\ket{\phi}}
        }
    \end{minipage}
\end{figure}

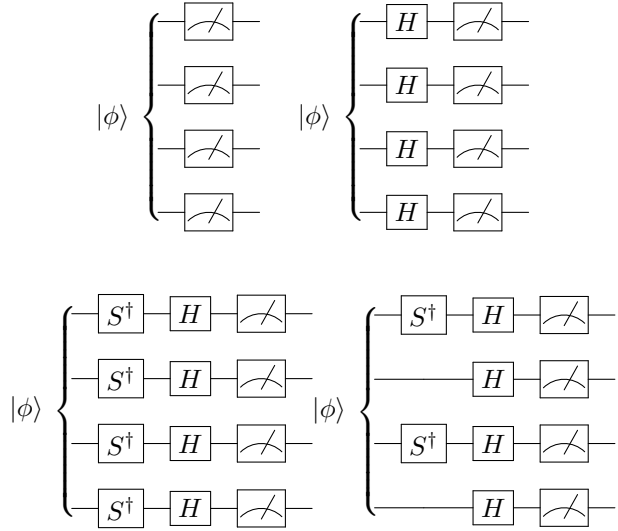
\begin{figure}[htbp]
    \centering
    \begin{minipage}[h]{0.45\linewidth}
        \Qcircuit @C=1em @R=1em {
            \lstick{} & \gate{S^\dagger} & \gate{H} & \meter & \qw \\
            \lstick{} & \gate{S^\dagger} & \gate{H} & \meter & \qw \\
            \lstick{} & \gate{S^\dagger} & \gate{H} & \meter & \qw \\
            \lstick{} & \gate{S^\dagger} & \gate{H} & \meter & \qw
            \inputgroupv{1}{4}{.5em}{3.6em}{\ket{\phi}}
        }
    \end{minipage}
    \hspace{0.5cm}
    \begin{minipage}[h]{0.45\linewidth}
        \Qcircuit @C=1em @R=1em {
            \lstick{} & \gate{S^\dagger} & \gate{H} & \meter & \qw \\
            \lstick{} & \qw              & \gate{H} & \meter & \qw \\
            \lstick{} & \gate{S^\dagger} & \gate{H} & \meter & \qw \\
            \lstick{} & \qw              & \gate{H} & \meter & \qw
            \inputgroupv{1}{4}{.5em}{3.6em}{\ket{\phi}}
        }
    \end{minipage}
    \caption{Four types of measurements on 4 qubit state $\ket{\phi}$.}
    \label{fig:ex_tom}
\end{figure}

Quantum state tomography is a method to estimate the amplitudes of quantum states.
An example of quantum state tomography for pure states \cite{2023PhRvA.107a2408V} is summarized as follows.

To estimate a $n$-qubit state $\ket{\phi}$, one performs four types of measurements are performed on $\ket{\phi}$.
The measurements for a 4-qubit state, for example, are shown in Fig. \ref{fig:ex_tom}.
Each measurement yields $d=2^n$ different outcomes from $00\cdots0$ to $11\cdots1$.
Because the cumulative distribution of the measurement obeys a multinomial distribution, the probability of obtaining each result with error $\epsilon$ can be estimated by performing the measurement $\mathcal{O}(d/\epsilon^2)$ times.
After obtaining the probabilities, an optimization problem is to be solved to maximize the likelihood. This can be achieved by using, for example, a method written in chapter 3 of \cite{paris2004quantum}.

\end{document}